\documentclass{aa}
\usepackage{graphicx}
\usepackage{txfonts}
\begin{document}

   \title{Identification of the broad solar emission features near 117 nm}

   \author{Eugene H. Avrett, Robert L. Kurucz, and Rudolf Loeser}

   \offprints{E. H. Avrett}

   \institute{Harvard-Smithsonian Center for Astrophysics, 
              60 Garden Street, Cambridge, MA 02138, USA \\
              \email{eavrett, rkurucz, rloeser@cfa.harvard.edu}}

   \date{Received 2005; accepted 2005}

  \abstract
{Wilhelm et al. have recently called attention to the unidentified broad
emission features near 117 nm in the solar spectrum.  They discuss the
observed properties of these features in detail but do not identify the 
source of this emission.  We show that the broad autoionizing 
transitions of neutral sulfur are responsible for these emission features.  
Autoionizing lines of \ion{S}{i} occur throughout the spectrum between Lyman alpha 
and the Lyman limit.  Sulfur is a normal contributor to stellar spectra. We use 
non-LTE chromospheric model calculations with line data from the
Kurucz 2004 \ion{S}{i} line list to simulate the solar spectrum in the range 116 
to 118 nm.  We compare the results with SUMER disk-center observations from Curdt et al.
and limb observations from Wilhelm et al.  Our calculations generally 
agree with the SUMER observations of the broad autoionizing \ion{S}{i} emission 
features, the narrow \ion{S}{i} emission lines, and the continuum in this wavelength
region, and agree with basic characteristics of the center-to-limb observations.
In addition to modeling the average spectrum, we show that a change of $\pm$ 200 K in the temperature
distribution causes the intensity to change by a factor of 4.  This
exceeds the observed intensity variations 1) with time in quiet regions at these 
wavelengths, and 2) with position from cell centers to bright network.  These results do 
not seem compatible with current dynamical models that have temporal variations 
of 1000 K or more in the low chromosphere.}
{}


\authorrunning{Avrett, Kurucz, \& Loeser}
\titlerunning{Broad solar emission features}
\maketitle
%

\section{Introduction}

This Letter is in response to the recent discussion by Wilhelm et al.
(\cite{wilhelm}) of emission features near 117 nm in the solar spectrum
that are much broader than observed emission lines in this wavelength range.  These authors rule out
a number of possible explanations, such as groups of emission lines blended
together, but they show that the center-to-limb behavior of these features
has a greater similarity to that of emission lines rather than to that of a
background continuum, and the results in their Fig. 3 and Table 1 show more 
similarity with \ion{S}{i} line emission than with line emission from \ion{C}{i}, \ion{He}{i}, or \ion{C}{iii}.

We identify these broad emission features as autoionizing transitions of
\ion{S}{i}.  Calculations indicate that there are 47 lines of \ion{S}{i} in the range
116 to 118 nm, of which 31 have distinct wavelengths (i.e., some transitions 
share the same wavelength).  All of these transitions
are between known energy levels and have upper energy levels located above
the \ion{S}{i} ionization threshold.  Six of the 31 are broad autoionization lines.  Four of
these six autoionization lines are at the wavelengths of the observed broad emission features.  The
remaining two are hidden by much stronger \ion{C}{iii} emission near 117.57 nm.

The autoionization lines are broad because they strongly interact with the ground state of \ion{S}{ii},
while the narrow \ion{S}{i} lines do not autoionize because their upper
levels can only weakly ionize to higher $^{2}P$ or $^{2}D$ states of \ion{S}{ii}.

Beyond identifying this wavelength correspondence, we calculate
the solar spectrum in the 116 to 118 nm wavelength range, using a 1-dimensional
model of the average quiet-Sun chromospheric temperature distribution that is
generally consistent with SUMER continuum observations, and with millimeter observations.
We find that we can account for the basic observed features reasonably well,
both at disk center and at the limb.

In Sect. V we show the intensities calculated from a model of the low
chromosphere with temperatures 200 K hotter than our average model and the
intensities from a model with temperatures 200 K cooler, and show that
this range of calculated intensities exceeds the range of both temporal
and spatial variations observed at these wavelengths on the quiet Sun.

\section{Line Data}
In 2004 Kurucz computed a line list, including line strengths, for \ion{S}{i} using methods described by 
Kurucz (\cite{kurucz}).  The line list and further details are given in the website 
http://kurucz.harvard.edu/atoms/1600. 

His calculation used 2161 even levels in 61 configurations and 2270 odd
levels in 61 odd configurations up to n = 16, resulting in 225605
electric dipole lines.  Of these, 24722 lines in file GF1600.pos are
between known energy levels and have good wavelengths.  The rest of the
lines have predicted wavelengths.  Once the lines were computed, the
widths and asymmetries of the autoionizing lines were adjusted using
Shore parameters (Shore \cite{shore}) to approximately match the 
observed autoionizing spectrum of Gibson et al. (\cite{gibson}).  
The calculations by Chen \& Robicheaux (\cite{chen}) and by Altun (\cite{altun})
served as a guide to understand the overlapping details and the background
photoionization continuum.

The adjusted autoionizing lines are in file GF1600.auto in the website.
The radiative, Stark, and van der Waals damping constants for the
autoionizing lines have been replaced by: the FWHM $\Gamma_{\rm{Shore}}$,
the asymmetry parameter $A_{\rm{Shore}}$, and the maximum cross-section
$B_{\rm{Shore}}$.  The profile is given by $A_{\rm{Shore}}\epsilon + B_{\rm{Shore}}/(\epsilon^2+1)$
where $\epsilon = 2(\nu - \nu_{\rm{line}})/\Gamma_{\rm{Shore}}$.

The data in the file GF1600.auto was substituted into GF1600.pos to make the file
GF1600.sub which contains both autoionizing and non-autoionizing lines.
The details for the 47 \ion{S}{i} lines between 116 and 118 nm, along with all
other \ion{S}{i} lines, can be found in that file.

An earlier calculation by Fawcett (\cite{fawcett}) included only 4 configurations
instead of the 122 used by Kurucz.  There are differences
between the Fawcett and Kurucz {\it gf} values of up to a factor of 2, probably caused by missing
configuration interactions in the Fawcett calculations.

\section{The computed and observed spectrum}

\begin{figure}
   \centering
\includegraphics[width=7cm, angle=90]{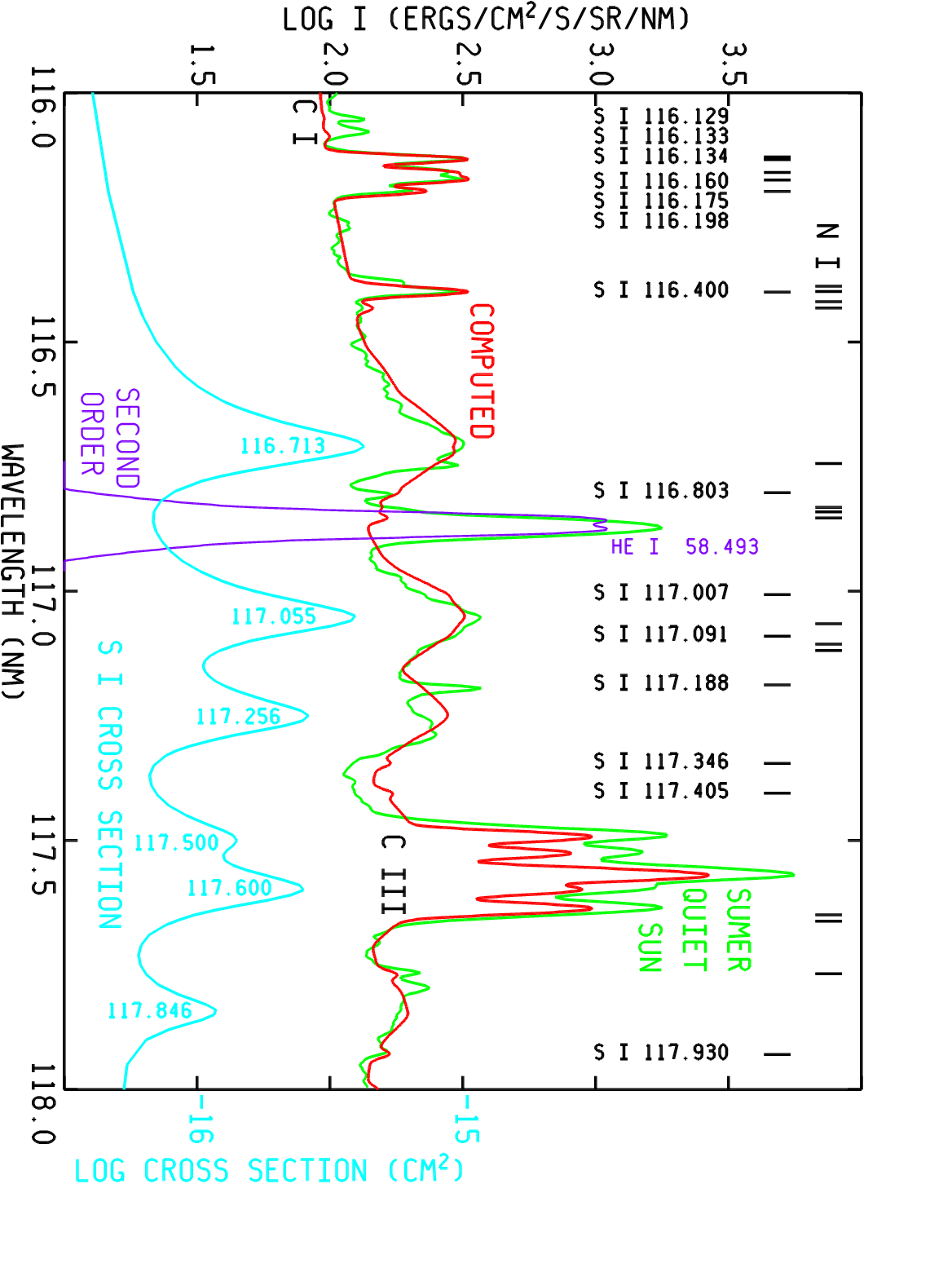}
   \caption{The upper curves show how the calculated disk-center intensity compares
with the SUMER observations in the 116 to 118 nm range.  The scale on the left gives the intensity, 
i.e., the spectral radiance.  The lower curve is the \ion{S}{i} photoionization cross section (right scale)
            that includes the six autoionization lines.}
              \label{FigSpec}%
    \end{figure}

Fig.~\ref{FigSpec} shows the observed disk-center intensity distribution for the
average quiet Sun between 116 and 118 nm from the SUMER atlas of Curdt et
al. (\cite{curdt}) together with our calculated intensities in the same absolute
units (scale on the left).  The lower part of this figure shows the \ion{S}{i} photoionization
cross section (scale on the right) which displays the six broad
autoionization lines.  There is a close match in wavelength and shape
between the calculated autoionization lines at 116.713 and 117.055 nm and the observed
broad emission features, and rough agreement at 117.256 and 117.846 nm.
The lines at 117.500 and 117.600 nm are blended with \ion{C}{iii} emission.

We include in this figure the 
calculated profiles of 14 narrow \ion{S}{i} emission lines having values of log {\it gf} 
larger than --3.0.  (Those at 117.007, 117.091, and 117.188 nm are too weak to 
be apparent.)  The strongest of these emission lines, at 116.198 nm, has log {\it gf} = 
--0.906, compared with --0.603 for the autoionization line at 116.713 nm.  The broad 
\ion{S}{i} autoionization lines contribute to the \ion{S}{i} continuum.  The total
continuum also includes non-LTE contributions from hydrogen, carbon, silicon, and other atoms. 
In establishing the non-LTE populations of the \ion{S}{i} levels we treat the autoionization 
lines only as part of the \ion{S}{i} photoionization cross section and do not include these lines 
as explicit transitions.

Figure 1 also shows the emission due to the strong \ion{C}{iii} multiplet with six
component lines between 117.493 and 117.637 nm, two \ion{C}{i} lines 
at 116.051 and 116.088 nm, and the \ion{He}{i} 58.433 nm resonance line from
the second-order spectrum which overlaps this mainly first-order spectrum at
116.867 nm.  We took this \ion{He}{i} line from our calculated spectrum near 58.4 nm, 
multiplied the wavelengths by 2 and the intensities by the factor 0.062 (determined from
the 1st and 2nd order scales shown in Fig. 4 of Curdt et al.) and added the result to Fig. 1.
The observed \ion{He}{i} line has a Gaussian shape while our calculated line has a central 
reversal.  We are able to diminish or eliminate this reversal by introducing flow velocities 
in the transition region where the line center is formed.

While the \ion{C}{iii} and \ion{He}{i} lines are formed much higher in the atmosphere 
than the other emission lines in this wavelength range,  we use the \ion{C}{iii} 
multiplet as follows to determine the broadening of our calculated line profiles to
compare with these observations.  In order to match the observed blending of the \ion{C}{iii} 
components, caused by solar atmospheric motions as well as by instrumental broadening in these 
observations, the calculated spectrum was convolved with a gaussian profile function having a FWHM of 0.015 nm.  
Lesser broadening applied to our calculated spectrum would give a separate emission peak just 
longward of the bright central peak, contrary to the observed partial blending of these 
two component lines.


We also include the 14 lines of \ion{N}{i} in this wavelength range having log {\it gf} $\ge$ --3.0. 
The wavelength positions of the \ion{N}{i} lines, along with those of \ion{S}{i}, are 
indicated at the top of the figure.  The four strongest \ion{N}{i} lines are at 
116.745, 116.854, 116.389, and 116.433 nm, with log {\it gf} values --0.675, --0.817, 
--1.038, and --1.249, respectively, according to the tables of Wiese et al. (\cite{wiese}).  The first of these appears on the red side of the 
116.713 nm broad emission feature.  The second is obscured by the overlapping second-order 
\ion{He}{i} line.  The third appears in the blue wing of the \ion{S}{i} 116.400 nm emission line.  
The fourth \ion{N}{i} line appears in our calculated spectrum just longward of the \ion{S}{i} 116.400 nm line but 
is not present in the observed spectrum, despite having a strength 0.6 times that 
of the nearby \ion{N}{i} 116.389 nm line from the same multiplet.  We cannot explain this 
discrepancy,
but note that the stronger line is blended with the \ion{S}{i} line while the weaker one
is not. This could be resolved when we have included the effects of blending between \ion{S}{i} and \ion{N}{i} lines.

We note that while the nitrogen abundance is almost 10 times that of sulfur, 
the \ion{N}{i} lines in this wavelength range are much weaker because their lower energy 
level is an excited state, and the \ion{N}{i} departures from LTE are greater than those 
of \ion{S}{i}. 
 
The narrow emission feature at
117.195 nm is an unidentified line (W. Curdt, private communication), and is
not the S I 117.188 nm line, which is too weak to appear. We defer 
a discussion of other lines and continua to a subsequent paper
(Avrett, Fontenla, \& Loeser, in preparation, hereafter AFL).

\section{Formation of the \ion{S}{i} spectrum}

We solve the coupled statistical equilibrium and radiative transfer
equations for a \ion{S}{i} atomic model with 23 energy levels and 111 line
transitions.  We must specify line strengths, line broadening
parameters, collisional excitation and ionization rates,
and photoionization cross sections.  Few of these values are
well known, but we can often determine through experimentation
which parameters critically affect the results and which
do not.  For example, the photoionization cross sections
and the collisional ionization rates for the lowest \ion{S}{i} levels
largely control the \ion{S}{i} contribution to the continuum
shortward of 120 nm.  Emission line strengths are generally 
sensitive to collisional excitation rates.  The \ion{S}{i} lines
considered here all have upper levels above the \ion{S}{ii} threshold
and are sensitive to the collisional coupling between these
levels and the \ion{S}{ii} continuum.

The photoionization rates depend on integrations over a large wavelength range that includes a large number of 
lines, some strongly in emission.  We cannot calculate the spectrum in the 116 to 118 nm region without 
solving the statistical equilibrium and radiative transfer equations
for hydrogen, \ion{Si}{i}, \ion{C}{i}, and several other constituents in
addition to \ion{S}{i}.  The AFL paper cited above will show results
of non-LTE calculations that include H, $\rm{H}^-$, \ion{He}{i-ii}, \ion{C}{i-iv}, \ion{N}{i-iv}, \ion{O}{i-vi}, 
\ion{Ne}{i-viii}, \ion{Na}{i-ii}, \ion{Mg}{i-ii}, \ion{Al}{i-ii}, \ion{Si}{i-iv}, \ion{S}{i-iv}, \ion{Ca}{i-ii}, 
and \ion{Fe}{i-ii}, applied to the interpretation of the SUMER atlas of 
Curdt et al. between 67 and 161 nm.  Our non-LTE
atmospheric modeling calculations use the Pandora computer
program of Avrett \& Loeser (\cite{avrett}).

Our current working model of the low chromosphere is listed in
Table~1.  This model is similar to the average quiet Sun model C 
of Vernazza et al. (\cite{vernazza}), updated by 
Fontenla et al. (\cite{fontenla}), but 
we have made adjustments to improve agreement with the observed 
distribution of brightness temperatures with wavelength in the 
millimeter range (see Loukitcheva et al. \cite{louk}) and with the continuum intensities in the SUMER 
atlas, including the continuum intensities shown here.  This model
will be revised further and presented in detail in the AFL paper
based on more complete comparisons with SUMER observations.

The table lists, as functions of height (above $\tau_{500\rm{nm}} = 1$), 
the adopted values of the temperature and a broadening velocity $V$.  
This broadening, or microturbulent, velocity is inferred from 
observed non-thermal doppler widths of lines formed at various
heights, and is used not only for line broadening but also as
a turbulent pressure velocity in determining the total hydrogen
number density $N_{\rm{H}}$ from hydrostatic equilibrium.  
The electron number density $N_{\rm{e}}$ is determined from the
degree of ionization of the various elements in the calculation.

\begin{table}
\caption{Adopted low chromospheric model for the average quiet Sun}
\label{table:1}      
\centering                          
\begin{tabular}{c c c c c}        
\hline\hline                 
Height(km) & $T$(K) & $V$(km $\rm{s}^{-1}$) & $N_{\rm{H}}(\rm{cm}^{-3})$ & $N_{\rm{e}}(\rm{cm}^{-3})$ \\
\hline                        
 1750  & 6785  &  8.17  &  4.17E+11  &  8.38E+10 \\
 1660  & 6750  &  7.64  &  6.19E+11  &  7.62E+10 \\
 1580  & 6710  &  7.16  &  8.90E+11  &  7.48E+10 \\
 1500  & 6670  &  6.66  &  1.30E+12  &  8.00E+10 \\
 1420  & 6615  &  6.14  &  1.94E+12  &  8.83E+10 \\
 1340  & 6540  &  5.58  &  2.97E+12  &  9.63E+10 \\
 1270  & 6420  &  5.06  &  4.45E+12  &  1.01E+11 \\
 1180  & 6200  &  4.32  &  7.87E+12  &  1.01E+11 \\
 1080  & 5840  &  3.37  &  1.62E+13  &  8.00E+10 \\
  990  & 5520  &  2.61  &  3.27E+13  &  6.35E+10 \\
  925  & 5290  &  2.27  &  5.56E+13  &  5.16E+10 \\
  860  & 5080  &  2.01  &  9.64E+13  &  4.37E+10 \\
  810  & 4940  &  1.77  &  1.50E+14  &  4.09E+10 \\
  770  & 4850  &  1.58  &  2.14E+14  &  4.22E+10 \\
  720  & 4750  &  1.35  &  3.38E+14  &  4.92E+10 \\
  660  & 4650  &  1.09  &  5.90E+14  &  6.85E+10 \\
\hline                                   
\end{tabular}
\end{table}
%

\section{Comparison of the broad emission and the continuum}

Consider the two wavelengths 116.316 and 116.713 nm.  The first is
a continuum wavelength relatively free of lines, and the second
is centered on the strongest broad emission feature.

\begin{figure}
   \centering
\includegraphics[width=9cm, angle=90]{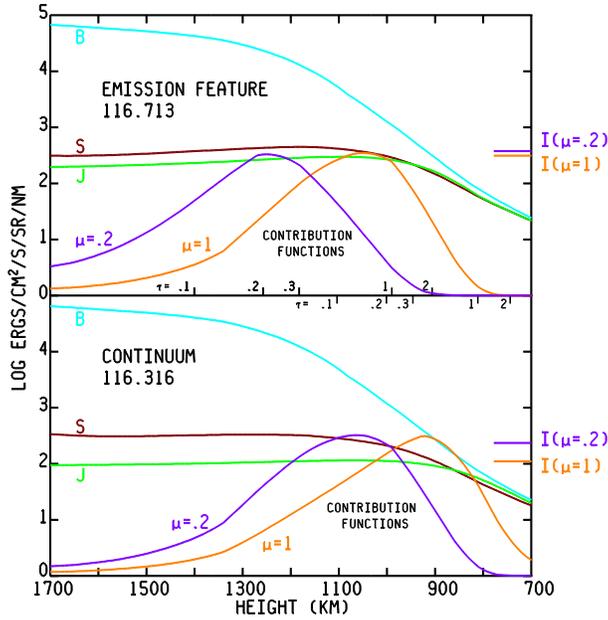}
   \caption{For the two wavelengths 116.316 and 116.713 nm: the variation with $h$ and with $\tau_{\lambda}$ 
            of $B$, $S$, and $J$, and of $dI/dh$ for $\mu = 1$ and $\mu = 0.2$. 
            The emergent intensity values for $\mu = 1$ and $\mu = 0.2$ are indicated on the right for each wavelength.}
              \label{FigSS}%
    \end{figure}

Fig.~\ref{FigSS} shows, for the two wavelengths, as functions of
both height $h$ and monochromatic optical depth $\tau_{\lambda}$: the Planck
function $B$ corresponding to the temperatures in Table 1, the continuum
source function $S$, the mean intensity $J$, and the function $dI/dh$
that gives the intensity contribution per unit height to the calculated
emergent intensity $I$ at disk center ($\mu = 1$) and near the limb
($\mu = 0.2$).  The two emergent intensity values are indicated on the right 
in both cases.  The emission feature has a larger intensity than the continuum 
because of its larger source funtion.

Since $S$ increases outwards, the formation region occurs at optical depths that are somewhat 
smaller than unity.  The two panels show that the peak intensity contributions occur 
higher in the atmosphere than $\tau = 1$ for the disk-center intensity,
or $\tau = 0.2$ for the limb intensity at $\mu = 0.2$. 

\begin{figure}
   \centering
\includegraphics[width=9cm, angle=90]{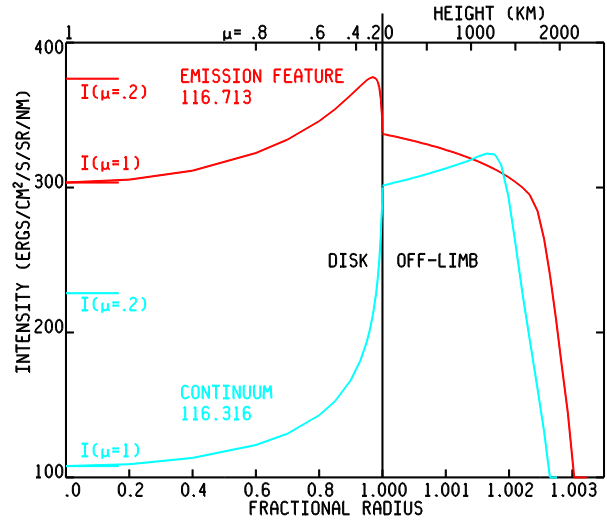}
      \caption{Calculated center-to-limb and off-limb intensity 
               variations at the two wavelengths.}
         \label{FigCLimb}
   \end{figure}

In Fig.~\ref{FigCLimb} we show the calculated center-to-limb and off-limb
intensities for the two wavelengths.  Both show brightening toward the limb, 
resulting from the outwardly increasing temperature.  The
intensity in the emission feature has a maximum just at, or inside, the limb, 
corresponding to the maximum of $S$ in the upper panel of Fig. 2.  
The continuum intensity is smaller on the disk, and exhibits a sharp 
increase just at the limb.  Above the limb, the calculated intensity in the emission feature
extends to greater heights than does the continuum intensity.  This is expected, 
since the opacity in the emission feature is greater than in the continuum.

The limb observations shown by Wilhelm et al. in their Fig. 5 show 
brightening toward the limb at both wavelengths, with continuum intensities much smaller than 
those in the emission feature,  and a maximum intensity in the emission
feature that appears to occur just inside the limb.  

The observed intensity in the emission
feature above the limb, however, extends further than in our calculation.  Their Fig. 6 shows that the
separation between the peak intensity of the emission feature
just inside the limb and 0.1 times this peak intensity above the limb
is about 5'' of arc, or about 3600 km, which is much larger
than the 2000 km or so in our calculation.  This can be interpreted as due to
the irregularities in height of the atmospheric layers, which are not accounted for by the 
assumed spherical symmetry in our calculations.


Our calculations also show that the continuum intensity reaches a maximum
at a very short distance above the limb, with a maximum value just above the
intensity of the emission feature at that location.  Their Fig. 5 shows extended 
patches of continuum emission above the limb that could be due to emission from extended 
inhomogeneous structures that are not represented in our 1-dimensional modeling.  The
results in their Fig. 8 and Table 2 indicate that the continuum intensity reaches a maximum 
just above the limb, as our calculations suggest, but that the continuum emission appears to remain
above that of the emission feature at greater heights, contrary to our results.  However,
Wilhelm (private communication) points out that the SUMER data available in December 2004 did not 
allow the photospheric limb position to be determined to the accuracy best suited for this comparision.

We have found that the calculated continuum intensity above the limb in this wavelength region 
is quite sensitive to the departures from LTE in H, \ion{S}{i}, \ion{Si}{i}, and \ion{C}{i}
in the higher layers of the atmosphere, i.e., to the effective scattering that we calculate in these 
layers due to the large number of emission lines overlapping the corresponding continua.  We 
continue to study these effects, particularly the relative importance of scattering and absorption 
in the many emission lines that affect these photoionization rates.  Further center-to-limb
observations at various EUV wavelenghts would be very useful for such studies.

\section{Temperature Variations}

\begin{figure}
   \centering
\includegraphics[width=7.5cm, angle=90]{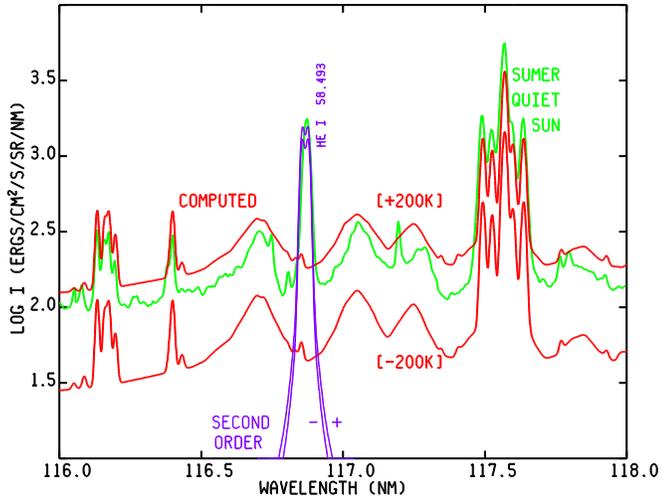}
   \caption{Intensities calculated from a model with chromospheric temperatures
            200 K higher and 200 K lower than in Table 1, 
            compared with the SUMER observations of the average quiet Sun.}
              \label{FigPlus}%
    \end{figure}

Finally we show how the calculated intensities are affected by
higher and lower model temperatures.  We calculate models based on
chromospheric temperatures 200 K higher and 200 K lower than in the model
listed in Table 1.  The two calculated intensity distributions in the 116 to 118 nm
range are shown in Fig.~\ref{FigPlus} along
with the same average quiet Sun observations as in Fig. 1.  The 200 K changes were 
introduced at all heights from the temperature minimum region into the transition region.  
From hydrostatic equilibrium, increasing/decreasing the temperature in the minimum region 
causes a smaller/larger decrease of $N_{\rm{H}}$ with height in the chromosphere, so that
the results in Fig.~\ref{FigPlus} reflect changes in density as well as temperature.
The ratio of intensities corresponding to the temperature changes between --200 K and +200 K 
is about 4 at the continuum wavelength 116.316 nm, and is about 3 for the emission feature 
at 116.713 nm.


The SUMER atlas of Curdt et al. gives the observed ratio of
bright network to cell interior as a function of wavelength.  This ratio
appears to be about 1.7 at the center of the strongest autoionization
feature, and about 2.0 at the continuum wavelength 116.316 nm.  Wilhelm et
al. show, in their Fig. 4, the observed variation with time (mainly
3-minute oscillations) of the autoionization emission.  The
intensity has peak-to-peak variations of about 1.5 in the inter-network, and
about 1.3 in network regions. 

The calculated range of intensities coresponding to $\pm$ 200 K is much greater than the range of observed
quiet-Sun intensity variations: 1) with time, as shown in Fig. 4 of Wilhelm et al.,
and 2) with position between inter-network and network regions, as shown
in the SUMER atlas.

These results show the temperature sensitivity of the calculated intensities.  
We do not claim to have determined the temperature distribution to within $\pm$ 200 K, 
since changing some of the important rates and cross sections, and changing the treatment 
of emission lines involved in photoionization, can have large effects.

\section{Conclusions}

We have demonstrated that the broad emission features discussed by Wilhelm
et al. are the result of \ion{S}{i} autoionization transitions.  Using
a 1-dimensional, time-independent model to represent the chromosphere
of the average quiet Sun, we have calculated the spectrum in the
116 to 118 nm band and have shown that the results roughly agree with the
disk-center quiet-Sun observations from the SUMER atlas of Curdt
et al.  Also, our calculated center-to-limb variations are similar to
the variations observed by Wilhelm et al.  We find that temperature variations of 
$\pm$ 200 K lead to calculated intensity variations much greater than the temporal 
and spatial variations observed in quiet solar regions.

With regard to such temperature variations, we note that Carlsson \& Stein 
(\cite{carlssonstein}) regard the low
chromosphere in Table 1 as wholly dynamic in nature, with temperatures
varying by 1000 K or more as individual shocks travel through the
atmosphere, and with significant time intervals during which the temperature has no
outward increase.  The 3D simulations of Wedemeyer et al. (\cite{wedemeyer})
support this view that the observed chromospheric emission does not necessarily imply 
an outward increase in the average gas temperature but can be explained by the presence of 
substantial spatial and temporal temperature inhomogenities.

However, as pointed out by Carlsson, Judge, \& Wilhelm (\cite{carlsson}),
simulations that do not have a persistent chromospheric temperature rise do
not qualitatively reproduce the behavior of the chromospheric emission lines
which are observed to be in emission at all times and all locations. A recent
discussion of this issue is provided by Fossum \& Carlsson \cite{fossum}.  

Given the variation of intensity with temperature shown here, and given
the moderate observed intensity fluctuations, both in continua and in lines, 
we conclude that models having a persistent outward temperature increase, 
but with moderate temporal and spatial variations, match observations better 
than current dynamical models that exhibit very large temperature fluctuations. 
Results for other wavelengths supporting this conclusion will be given in subsequent papers.

\begin{acknowledgements}
We thank Juan Fontenla, Adriaan Van Ballegooijen, and the referee, Klaus Wilhelm, for their comments.
\end{acknowledgements}

\end{document}